\documentclass[aps,prd,preprint,superscriptaddress,amsmath,amssymb,showpacs]{revtex4-1}
\usepackage{dcolumn}
\usepackage{graphicx}
\usepackage{float}
\usepackage{physics}
\usepackage[colorlinks=true,allcolors=blue]{hyperref}
\usepackage{amsmath}
\usepackage{color}
\begin{document}

\title{Moving doubly heavy baryon in a strongly coupled plasma from holography}

\author{Xuan Liu}
\affiliation{School of Nuclear Science and Technology, University of South China, Hengyang 421001, China}
\author{Jia-Jie Jiang}
\affiliation{School of Nuclear Science and Technology, University of South China, Hengyang 421001, China}
\author{Xun Chen}
\email{chenxunhep@qq.com}
\affiliation{School of Nuclear Science and Technology, University of South China, Hengyang 421001, China}
\author{Mitsutoshi Fujita}
\email{fujitamitsutoshi@usc.edu.cn}
\affiliation{School of Nuclear Science and Technology, University of South China, Hengyang 421001, China}
\author{Akira Watanabe}
\email{watanabe@usc.edu.cn}
\affiliation{School of Mathematics and Physics, University of South China, Hengyang 421001, China}

\begin{abstract}
    Gauge/gravity duality is used to study properties of the doubly heavy baryon $(\mathrm{QQq})$ at finite rapidity and temperature in the heavy-ion collision. We investigate the impact of rapidity on string breaking and screening of $\mathrm{QQq}$, and compare these effects with the results for $\rm Q\bar{Q}$ in detail. Computations reveal that the string-breaking distances of $\mathrm{QQq}$ and $\rm Q\bar{Q}$ are close in the confined state and the effects of rapidity and temperature on the string breaking are not significant. An interesting result shows that $\mathrm{QQq}$ cannot be found at high enough temperatures and rapidities; however, $\rm Q\bar{Q}$ can exist under any conditions as long as the separation distance is small enough. Besides, the screening distances of $\mathrm{QQq}$ and $\rm Q\bar{Q}$ are also compared at finite rapidity and temperature. Based on the above analysis, we infer that $\rm Q\bar{Q}$ is more stable than $\mathrm{QQq}$ at finite rapidity and temperature.

\end{abstract}

\maketitle
\flushbottom

\section{Introduction}\label{sec:01_intro}

In the recent $\mathrm{LHCb}$ experiment at $\mathrm{CERN}$, researchers have made an exciting discovery of a particle known as $\Xi_{cc}^{++}$ \cite{LHCb:2017iph, LHCb:2018pcs}. This groundbreaking finding has sparked significant interest in the exploration of double charm baryons. The $\Xi_{cc}^{++}$ particle is formed by the combination of two heavy quarks with a single light quark, resulting in a highly distinctive structure. In this study, we adopt an assumption of potential models, which suggests that the interaction between heavy quarks and anti-quarks within hadrons can be described by a potential energy \cite{Eichten:1979ms}.

While lattice gauge theory remains a fundamental tool for studying non-perturbative phenomena in Quantum Chromodynamics $\mathrm{(QCD)}$, its investigation of the doubly heavy baryon potential has been relatively limited so far \cite{Yamamoto:2008jz, Najjar:2009da}. On the other hand, gauge/gravity duality provides a new theoretical tool for studying strongly coupled gauge theories, and in Ref. \cite{Maldacena:1998im}, the holographic potential of quark-antiquark pairs was computed for the first time. Based on the study in Ref. \cite{Maldacena:1998im}, the potential was further investigated in Refs. \cite{Rey:1998bq, Brandhuber:1998bs} to study the potential at finite temperature. In recent years, the methodology of utilizing holographic theories to study multi-quark potentials has been increasingly refined. The obtained multi-quark potentials from holography are in good agreement with lattice calculations and $\mathrm{QCD}$ phenomenology \cite{Alexandrou:2001yt, Alexandrou:2002sn, Takahashi:2002bw, Andreev:2020xor, Andreev:2015iaa, Andreev:2015riv, Andreev:2019cbc, Andreev:2021bfg, Andreev:2022qdu, Andreev:2022cax, Andreev:2023hmh, Andreev:2021eyj}. The techniques employed to extract the $\mathrm{QQq}$ potential in our study are analogous to those used in lattice $\mathrm{QCD}$ \cite{Yamamoto:2008fm}.

The creation of quark-gluon plasma $\mathrm{(QGP)}$ through high-energy heavy-ion collisions allows us to simulate the high temperature and density conditions of the early universe \cite{Rothkopf:2019ipj, Hindmarsh:2005ix, Asaka:2006rw, Laine:2006we}. The formation of this $\mathrm{QGP}$ is crucial for our understanding of the early stages of cosmic evolution and the fundamental properties of quark color dynamics. At low temperatures, quarks and gluons are confined within hadrons. As the separation between quarks increases, the strong interaction force becomes stronger. When the separation reaches a limit, new quark-antiquark pairs are typically produced, resulting in the confinement of quarks within hadrons. This behavior is manifested as string breaking in this study. Under extreme conditions, the strong interaction between hadrons diminishes at long range, allowing quarks to move freely over larger distances\cite{Matsui:1986dk, Witten:1998zw, Kharzeev:1995kz}.

In heavy-ion collision experiments, heavy ions collide at speeds close to the speed of light, forming $\mathrm{QGP}$ \cite{Rothkopf:2019ipj}. Similar to the early universe, this substance does not remain stationary but rapidly expands. Therefore, when studying the doubly heavy baryon, the effect of rapidity must be considered. This is crucial for understanding the interaction between $\mathrm{QQq}$ and the medium, as well as the transport properties within the $\mathrm{QGP}$. By studying the behavior of moving $\mathrm{QQq}$, more comprehensive information about the $\mathrm{QGP}$ can be obtained, leading to further investigation into the properties of $\mathrm{QCD}$. By studying the variation of string breaking under different temperature and rapidity conditions, we can study decay rates of different hadrons in different conditions\cite{Bigazzi:2006pn}.

Gauge/gravity duality was initially proposed by Maldacena \cite{Maldacena:1997re} for conformal field theories, and later extended to include theories resembling $\mathrm{QCD}$, establishing a connection between string theory and heavy-ion collisions in some manner \cite{Aharony:1999ti, Gubser:2011qv, Casalderrey-Solana:2011dxg}. The study of moving heavy quarkonium can be found in Refs. \cite{Liu:2006nn, Chen:2020ath, Chen:2017lsf, Song:2007gm, Finazzo:2014rca, Ali-Akbari:2014vpa, Andreev:2021vjr, Krishnan:2008rs, Yao:2020eqy, Chernicoff:2012bu, Benzahra:2002we, Thakur:2016cki, BitaghsirFadafan:2015yng, Escobedo:2013tca, Zhou:2020ssi, Feng:2019boe, Zhou:2021sdy}.

The remaining sections of this paper are as follows: In Sec. \ref{sec:02}, we will construct string configurations considering the effects of finite temperature and finite rapidity. In Sec. \ref{sec:03}, we will discuss the computation of the model's potential energy. At first, we calculate the influence of rapidity on the string breaking of $\mathrm{QQq}$. Then, we discuss its screening and provide a comprehensive comparison with $\mathrm{Q\Bar{Q}}$. Additionally, we also plot their state diagram in the $T- \eta$ plane. Finally, we summarize these findings in Sec. \ref{sec:04}.

\section{Model setup}\label{sec:02}

First, the background metric at a finite temperature is given as \cite{Andreev:2015riv, Chen:2021bkc, Andreev:2015iaa, Andreev:2007rx, Andreev:2007zv}:
\begin{align}
    ds^{2}&=w(r) \Big (-f(r)dt^{2}+ d\Vec{x}^{2}  +\frac{1}{f(r)}dr^{2} \Big )+e^{-\mathbf{s}r^{2}}g_{ab}^{(5)}d\omega^{a}d\omega^{b},
\end{align}
where
\begin{equation}
    \begin{aligned}
        w(r)&=\frac{e^{\mathbf{s}r^{2}}R^{2}}{r^{2}}\,,\\ f(r)&=1-\frac{r^{4}}{r_{h}^{4}}.\label{2}
    \end{aligned}
\end{equation}
The metric represents a one-parameter deformation of Euclidean $\mathrm{AdS_5}\times\mathrm{S_5} $ space, parametrized by $\mathbf{s}$ and with a radius $R$. It consists of a five-dimensional compact space (sphere) $\mathrm {X}$ with coordinates $\omega^{a}$, a blackening factor $f(r)$, and a black hole horizon (brane) at $r_h$. The Hawking temperature of the black hole $T$ is defined as follows:
\begin{equation}
    T=\frac{1}{4\pi}{\left | \frac{df}{dr}  \right |}_{r=r_{h}}=\frac{1}{\pi r_{h}}.\label{3}
\end{equation}

In this work, we investigate the motion of a particle consisting of two heavy quarks and one light quark in a thermal medium. We divide the particle's spatial coordinates into three directions: $x_{1}$ (connecting the heavy quark pair), $x_{2}$, and $x_{3}$ (perpendicular to the heavy quark pair).  We find that the coefficients of $x_1$ and $x_2$ in the metric are the same, which implies that the presence of the heavy quark pair at either $x_1$ or $x_2$ does not affect the calculation results.  Due to the temperature and rapidity being constant, we can consider the doubly heavy baryon to be in a state of force equilibrium. Furthermore, energy loss has not been taken into account in this paper, which can be discussed in future work.

In our study, we consider where the $\mathrm{QQq}$ is moving with a constant rapidity $\eta$ along the $x_{3}$ direction \cite{Liu:2006nn, Finazzo:2014rca, Chen:2017lsf, Andreev:2021vjr, Thakur:2016cki}. For convenience, let us consider a new scenario in which the $\mathrm{QQq}$ is at rest, while the surrounding medium is in motion relative to it. We then introduce the Lorentz transform to reflect the effect of rapidity $\eta$ on it and obtain the new metric:
\begin{align}
    ds^{2}=w(r)&\Big(-g_{1}(r)dt^{2}
 -2\sinh(\eta)\cosh(\eta)\big(1-\frac{g_{1}(r)}{g_{2}(r)}\big)dx_{3}dt\notag\\ &+g_{3}(r)dx_{3}^{2}+dx_{1}^{2}+dx_{2}^{2}+\frac{g_{2}(r)}{g_{1}(r)}dr^{2}\Big)+e^{-\mathbf{s}r^{2}}g_{ab}^{(5)}d\omega^{a}d\omega^{b}, \label{4}
\end{align}
where
\begin{equation}
    \begin{aligned}
        g_{1}(r)&=f(r)\cosh^{2}(\eta)-\sinh^{2}(\eta),\\g_{2}(r)&=\frac{g_{1}(r)}{f(r)},\\g_{3}(r)&=\cosh^{2}(\eta)-f(r)\sinh^{2}(\eta).\label{5}
    \end{aligned}
\end{equation}

The Nambu-Goto action of a string is
\begin{equation}
     S_{NG} = - \frac{1}{2\pi\alpha'} \int d\xi^{0} d\xi^{1} \sqrt{- \det g_{ab}},
\end{equation}
where $\alpha'$ is a constant, ($\xi^{0}$, $\xi^{1}$) are worldsheet coordinates, and $g_{ab}$ is an induced metric.
We introduce the baryon vertex and heavy and light quarks to construct the $\mathrm{QQq}$ configuration. Then we know from $\mathrm{AdS/CFT}$ that
the vertex is a five brane \cite{Witten:1997bs, Gukov:1998kn}. At leading order in $\alpha'$, the baryon vertex action is $S_{vert}=\mathrm{\tau}_{5}\int d^{6}\xi \sqrt{\gamma^{(6)}}$, where $\tau_{5}$ is the brane tension and $\xi^{i}$ is the worldvolume coordinates. Since the brane is wrapped on the $\mathrm {X}$ interior space, it looks point-like in $\mathrm{AdS_5}$. We choose a static gauge $\xi^{0}=t,\,\xi^{a}=\theta^{a}$ with $\theta^{a}$ is the coordinates on $\mathrm {X}$. So, the action is
\begin{equation}
    S_{vert}=\tau_{v}\int dt\frac{e^{-2\mathbf{s}r^2}}{r} \sqrt{g_{1}(r)},
\end{equation}
where $\tau_{v}$ is a dimensionless parameter defined by $\tau_{v}=\mathrm{\tau}_{5}R\mathrm{vol(X)}$ and $\mathrm{vol(X)}$ is a volume of $\mathrm {X}$.

Finally, we consider the light quark at the end of the string as a tachyon field, which couples to the worldsheet boundary by $S_q=\int d\tau \mathbf{e}\mathrm{T}$ \cite{Andreev:2020pqy, Erlich:2005qh}. This term is typical for strings propagating in an open string tachyon background and $\mathrm{T}$ is its background scalar \cite{Andreev:2020pqy}. The integral is over a worldsheet boundary parameterized by $\tau$ and $\mathbf{e}$ represents the boundary metric. In this paper, we can assume $\mathrm{T}$ to be a constant $\mathrm{T}_{0}$, and the integral reduces to a coordinate integral of $t$. Thus, the action can be given as
\begin{equation}
    S_{q}=m\int dt\frac{e^{\frac{\mathbf{s}r^2}{2}}}{r} \sqrt{g_{1}(r)},
\end{equation}
where $m=R\mathrm{T}_{0}$. This represents a particle with a mass $\mathrm{T}_{0}$ is at rest, while the surrounding medium with a temperature $T$ moves relative to it with a rapidity $\eta$. The parameters we select are as follows: $g=\frac{R^2}{2\pi\alpha'}=0.176,\,k=\frac{\tau_{v}}{3g}=-0.321, n=\frac{m}{g}=3.057,\mathbf{s}=0.45\,\mathrm{GeV^2},\,c=0.623\,\mathrm{GeV}$ \cite{Andreev:2021bfg}.

\subsection{Small L}
\begin{figure}
    \centering
    \includegraphics[width=8.5cm]{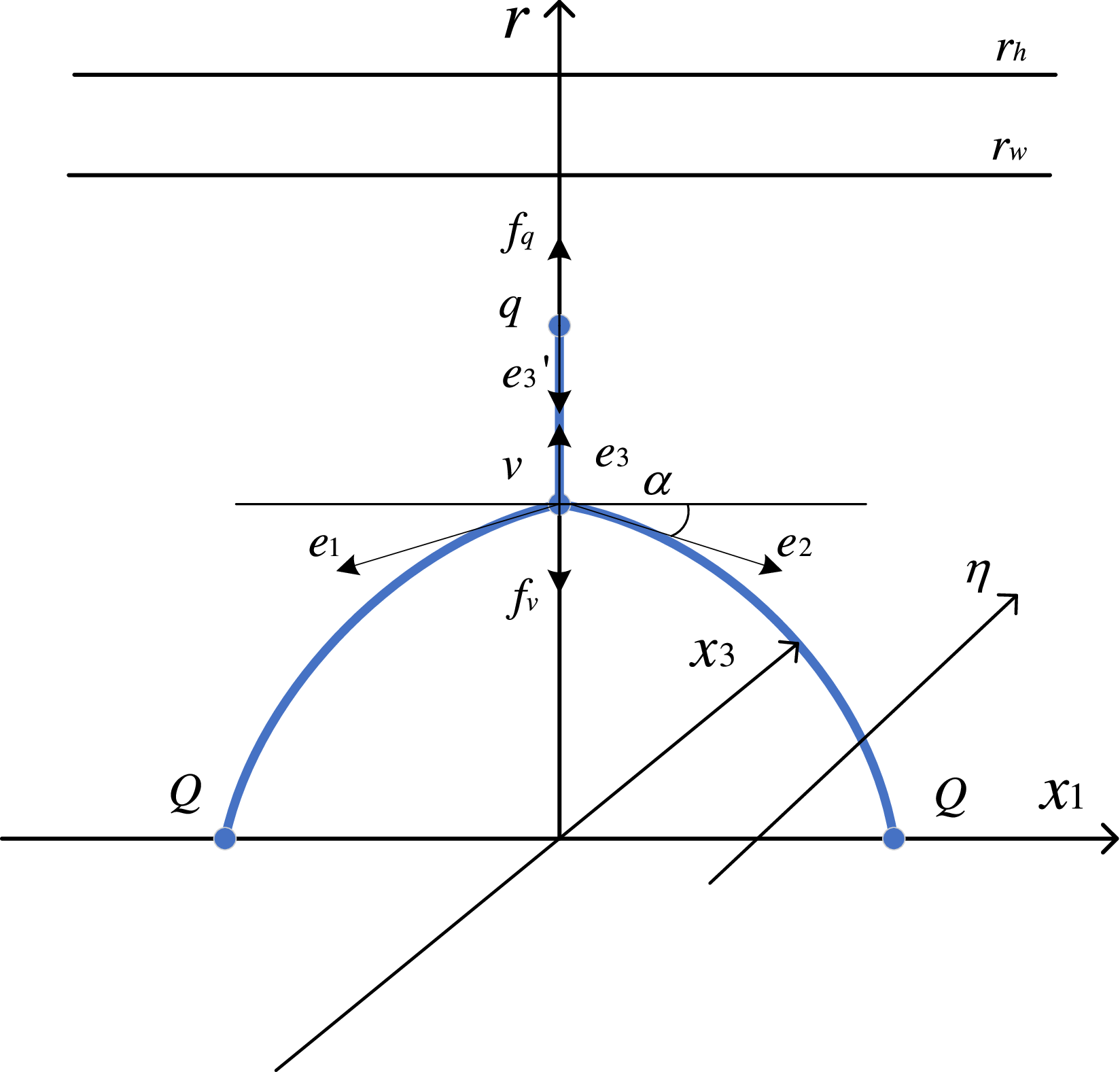}
    \caption{\label{fig1}  A string configuration at a small separation distance of heavy quark pair. We use the line connecting two heavy quarks $\mathrm{Q}$ as the $x_1$-axis, while the baryon vertex $\mathrm{v}$ and the light quark $\mathrm{q}$ are on the $r$-axis. The rapidity $\eta $ is along the $x_3$-axis direction. The heavy quark, light quark, and baryon vertex are connected by blue strings. The black arrows represent generalized forces. $r_{h}$ is the position of the black hole horizon. $r_{w}$ is the position of an imaginary wall when the $\mathrm{QQq}$ is confined.}
\end{figure}

As shown in Fig.~\ref{fig1}, the heavy quark pair, light quark, and baryon vertex are connected by three strings, with the separation distance between the heavy quark pair denoted as $L$. Therefore, we can express the total action as follows:
\begin{equation}
    S=\sum_{i=1}^{3} S_{NG}^{(i)}+S_{vert}+S_{q}.
\end{equation}
According to the previous discussion, since the results obtained from $x_{1}$ and $x_{2}$ are the same, we directly designate the line where the heavy quark pair is located as  $x_1$, and the rapidity perpendicular to the $x_1$ direction as $x_3$, as shown in Fig.~\ref{fig1}. Hereafter, we refer to $x_1$ as $x$. Then, in this configuration, we choose the static gauge where $\xi^{0}=t,\,\xi^{1}=r$, and consider $x$ as a function of $r$. Therefore, the total sum of the Nambu-Goto action can be written as:
\begin{equation}
    \sum_{i=1}^{3} S_{NG}^{(i)}=2g \int_{0}^{t} dt \int_{0}^{r_{v}} w(r)\sqrt{g_{1}(r)(\partial_{r}x)^2+g_{2}(r)} dr+g\int_{0}^{t} dt \int_{r_{v}}^{r_{q}} w(r)\sqrt{g_{1}(r)(\partial_{r}x)^2+g_{2}(r)}dr,
\end{equation}
where $\partial_{r}x=\frac{\partial x}{\partial r}$ and $x(r)$ satisfies the following boundary conditions:
\begin{equation}
    x(0)=\pm\,\frac{L}{2},\,x(r_{v})=x(r_{q})=0,\,
    \begin{cases}
    (\partial_{r}x)^2=\cot^2(\alpha),&r=r_{v}\\
      (\partial_{r}x)^2=0,&r\in (r_{v},r_{q}].
    \end{cases}
\end{equation}

Now, we combine the baryon vertex and the light quark to write that the total action is
\begin{align}
    S=gt&\Big(2\int_{0}^{r_{v}} w(r)\sqrt{g_{1}(r)(\partial_{r}x)^2+g_{2}(r)} dr+ \int_{r_{v}}^{r_{q}} w(r)\sqrt{g_{2}(r)}dr\notag\\
    &+3k\frac{e^{-2\mathbf{s}r^2}}{r} \sqrt{g_{1}(r)}+n\frac{e^{\frac{\mathbf{s}r^2}{2}}}{r} \sqrt{g_{1}(r)}\Big).
\end{align}
It is easy to find that the first term is divergent, so we get the potential energy according to $E=S/t$ after normalization
\begin{align}
    E_{QQq}=g&\Big(2\int_{0}^{r_{v}} \big(w(r)\sqrt{g_{1}(r)(\partial_{r}x)^2+g_{2}(r)}-\frac{1}{r^{2}}\big)dr-\frac{2}{r_{v}}+\int_{r_{v}}^{r_{q}} w(r)\sqrt{g_{2}(r)}dr\notag\\
    &+3k\frac{e^{-2\mathbf{s}r_{v}^2}}{r_{v}} \sqrt{g_{1}(r_{v})}+n\frac{e^{\frac{\mathbf{s}r_{q}^2}{2}}}{r_{q}} \sqrt{g_{1}(r_{q})}\Big)+2c.\label{13}
\end{align}
With $r_{v}$ as the independent variable at fixed $T$ and $\eta$.
By substituting the first term of the unnormalized energy into the Euler-Lagrange equation, we obtain:
\begin{equation}
    \mathcal{H}=\frac{w(r)g_{1}(r)\partial_{r}x}{\sqrt{g_{1}(r)(\partial_{r}x)^2+g_{2}(r)}},
\end{equation}
since $\mathcal{H}$ is a constant, we have:
\begin{equation}
    \mathcal{H}|_{r=r_{v}}=\frac{w(r_{v})g_{1}(r_{v})\cot (\alpha)}{\sqrt{g_{1}(r_{v})\cot^2(\alpha)+g_{2}(r_{v})}}.
\end{equation}
According to $\mathcal{H}=\mathcal{H}|_{r=r_{v}}$,
\begin{equation}
    \partial_{r}x=\sqrt{\frac{w(r_{v})^{2}g_{1}(r_{v})^{2}g_{2}(r)}{w(r)^{2}g_{1}(r)^{2}\big(g_{1}(r_{v})+g_{2}(r_{v})\tan^2(\alpha)\big)-g_{1}(r)w(r_{v})^{2}g_{1}(r_{v})^{2}}}.
\end{equation}

Now we need to determine the position of the light quark $r_{q}$ and the angle $\alpha$ between the string and the $x$-axis at the baryon vertex. The generalized force equilibrium equations at the baryon vertex and the light quark must be equal to zero. First, we take the variation of their respective action quantities to obtain their generalized forces:
\begin{align}
    e_{1}&=gw(r_{v})\Big(-\sqrt{\frac{g_{1}(r_{v})f(r_{v})}{f(r_{v})+\tan^{2}\alpha}},\,-\sqrt{\frac{g_{1}(r_{v})}{f(r_{v})^2\cot^2(\alpha)+f(r_{v})}}\Big),\\
    e_{2}&=gw(r_{v})\Big(\sqrt{\frac{g_{1}(r_{v})f(r_{v})}{f(r_{v})+\tan^{2}\alpha}},\,-\sqrt{\frac{g_{1}(r_{v})}{f(r_{v})^2\cot^2(\alpha)+f(r_{v})}}\Big),\\
    e_{3}&=gw(r_{v})\big(0,\,\sqrt{g_{2}(r_{v})}\big),\label{19}\\
    e_{3}'&=gw(r_{q})\big(0,\,-\sqrt{g_{2}(r_{q})}\big),\\
    f_{q}&=\Big(0,\,-gn\partial_{r_{q}}\big(\frac{e^{\frac{\mathbf{s}r_{q}^2}{2}}}{r_{q}} \sqrt{g_{1}(r_{q})}\big)\Big),\\
    f_{v}&=\Big(0,\,-3gk\partial_{r_{v}}\big(\frac{e^{-2\mathbf{s}r_{v}^2}}{r_{v}} \sqrt{g_{1}(r_{v})}\big)\Big),\label{22}
\end{align}
where $e_{i}$ is the string tension, and $f_{q}$ and $f_{v}$ are the forces provided by the light quark and baryon vertices, respectively \cite{Andreev:2021bfg}.

It is evident that the generalized force equilibrium equation at the light quark is solely a function of $r_{q}$ and can be expressed as follows:
\begin{equation}
    F_{1}(r_{q})=\frac{f_{q}+e_{3}'}{gw(r_{q})}=0.\label{23}
\end{equation}
In addition, we can also solve for $r_{q}$ based on this equation by changing $T$ and $\eta$. The force balance equation at the vertex is:
\begin{equation}
    f_{v}+e_{3}+e_{1}+e_{2}=0.\label{24}
\end{equation}
In addition to $T$ and $\eta$, the equation has only two unknowns, $r_{v}$, and $\alpha$, By solving the equation, we can obtain the function of $\alpha$ as a function of $r_{v}$.
Then we can give the distance $L$ as a function of $r_{v}$:
\begin{equation}
    L=2\int_{0}^{r_{v}}\frac{\partial x}{\partial r}  dr.\label{25}
\end{equation}
Together with Eqs. (\ref{13}) and (\ref{25}), we can numerically solve the energy at small $L$.

\subsection{Intermediate L}
\begin{figure}
    \centering
    \includegraphics[width=8.5cm]{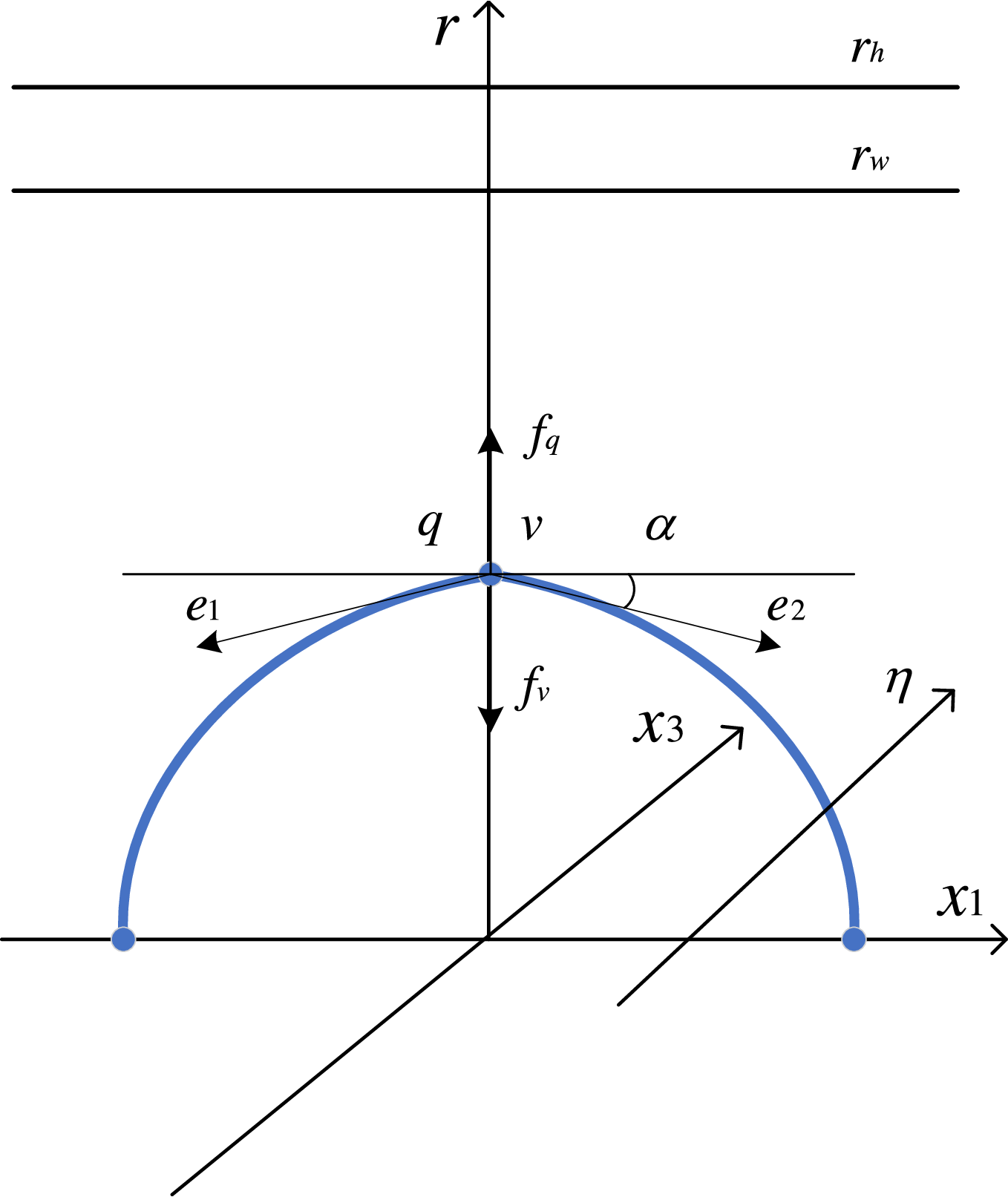}
    \caption{\label{fig2} A string configuration at an intermediate separation distance of a heavy quark pair. We use the straight line between two heavy quarks $\mathrm{Q}$ as the $x_1$-axis, and baryon vertex and light quarks are in the same position on the $r$-axis. The rapidity $\eta $ is along the $x_3$-axis direction. The black arrows represent generalized forces. $r_{h}$ is the position of the black hole horizon. $r_{w}$ is the position of an imaginary wall when the $\mathrm{QQq}$ is confined.}
\end{figure}

According to Fig.~\ref{fig2}, compared to small $L$, intermediate $L$ is only missing a straight string from $r_{v}$ to $r_{q}$, and the sum of the action can be written as
\begin{equation}
    S=\sum_{i=1}^{2} S_{NG}^{(i)}+S_{vert}+S_{q}.
\end{equation}
We still chose the same static gauge as before. It is easy to know that the sum of Nambu-Goto action becomes
\begin{equation}
    \sum_{i=1}^{2} S_{NG}^{(i)}=2g \int_{0}^{t} dt \int_{0}^{r_{v}} w(r)\sqrt{g_{1}(r)(\partial_{r}x)^2+g_{2}(r)} dr.
\end{equation}
Then the boundary condition becomes
\begin{equation}
    x(0)=\pm\,\frac{L}{2},\,x(r_{v})=0,\,(\partial_{r}x|_{r=r_{v}})^2=\cot^2(\alpha).
\end{equation}
Therefore, the potential energy is
\begin{align}
    E_{QQq}=g&\Big(2\int_{0}^{r_{v}} \big(w(r)\sqrt{g_{1}(r)(\partial_{r}x)^2+g_{2}(r)}-\frac{1}{r^{2}}\big)dr-\frac{2}{r_{v}}\notag\\
    &+3k\frac{e^{-2\mathbf{s}r_{v}^2}}{r_{v}} \sqrt{g_{1}(r_{v})}+n\frac{e^{\frac{\mathbf{s}r_{v}^2}{2}}}{r_{v}} \sqrt{g_{1}(r_{v})}\Big)+2c.\label{29}
\end{align}

Finally, we can write the equation for the equilibrium of generalized force at the coincidence of the light quark and the baryon vertex as
\begin{equation}
    f_{v}+f_{q}+e_{1}+e_{2}=0,\label{30}
\end{equation}
and each force is given by
\begin{align}
    e_{1}&=gw(r_{v})\Big(-\sqrt{\frac{g_{1}(r_{v})f(r_{v})}{f(r_{v})+\tan^{2}\alpha}},\,-\sqrt{\frac{g_{1}(r_{v})}{f(r_{v})^2\cot^2(\alpha)+f(r_{v})}}\Big),\\
    e_{2}&=gw(r_{v})\Big(\sqrt{\frac{g_{1}(r_{v})f(r_{v})}{f(r_{v})+\tan^{2}\alpha}},\,-\sqrt{\frac{g_{1}(r_{v})}{f(r_{v})^2\cot^2(\alpha)+f(r_{v})}}\Big),\\
    f_{q}&=\Big(0,\,-gn\partial_{r_{v}}\big(\frac{e^{\frac{\mathbf{s}r_{v}^2}{2}}}{r_{v}} \sqrt{g_{1}(r_{v})}\big)\Big),\\
    f_{v}&=\Big(0,-3gk\partial_{r_{v}}\big(\frac{e^{-2\mathbf{s}r_{v}^2}}{r_{v}} \sqrt{g_{1}(r_{v})}\big)\Big).
\end{align}
According to Eqs. (\ref{25}) and (\ref{29}), we can obtain the energy at intermediate $L$.

\subsection{Large L}

\begin{figure}
    \centering
    \includegraphics[width=8.5cm]{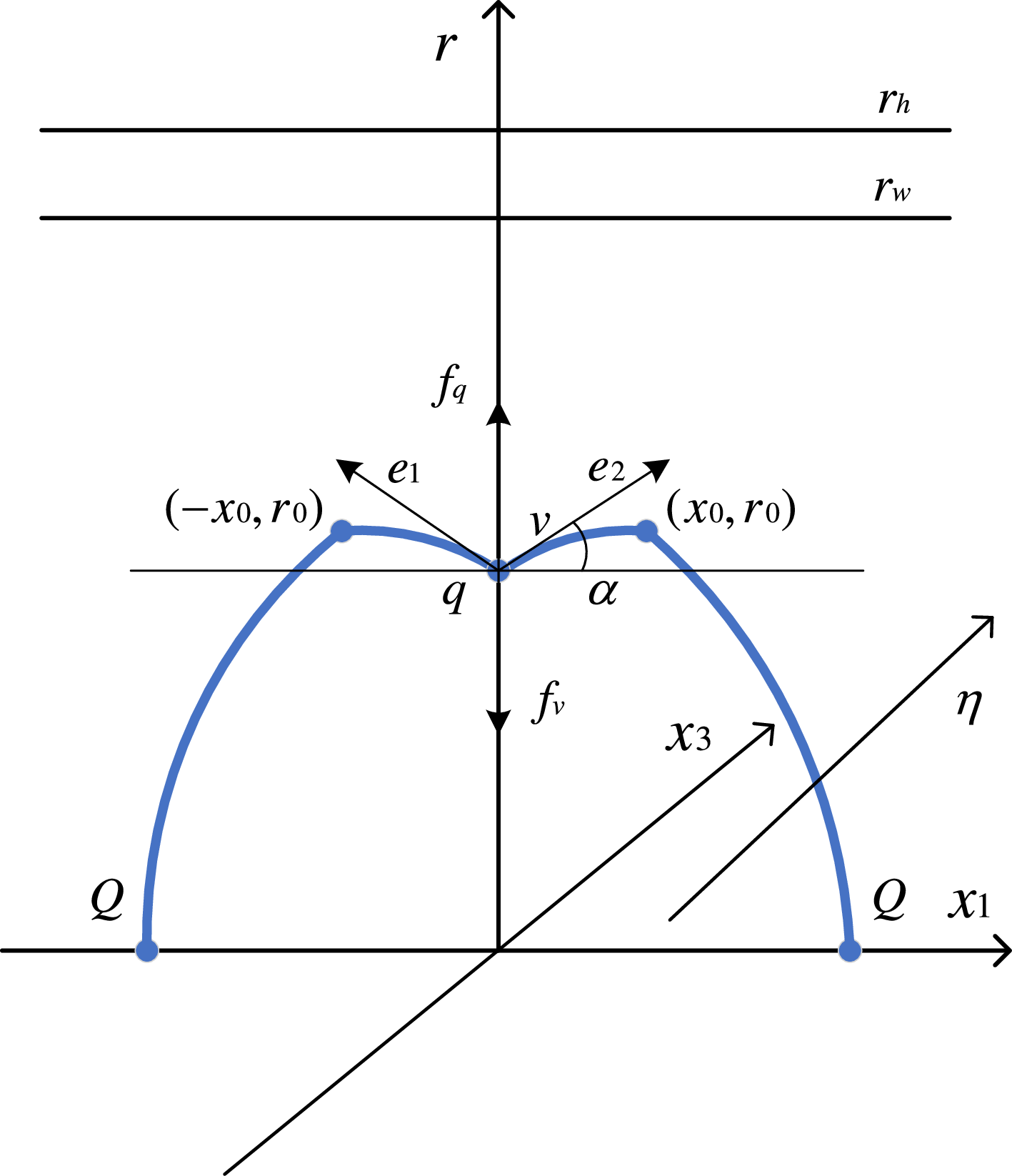}
    \caption{\label{fig3} A string configuration at a large separation distance of a heavy quark pair. We use the straight line between two heavy quarks $\mathrm{Q}$ as the $x_1$-axis, and baryon vertex and light quarks are in the same position on the $r$-axis. The rapidity $\eta $ is along the $x_3$-axis direction. There is a smooth turning point on each of the two strings. The black arrows represent generalized forces. $r_{h}$ is the position of the black hole horizon. $r_{w}$ is the position of an imaginary wall when the $\mathrm{QQq}$ is confined.}
\end{figure}

As can be seen from Fig.~\ref{fig3}, the configuration still consists of two strings, the vertex, and the light quark. However, due to the presence of smooth turning points on the strings, we can utilize a new static gauge that simplifies the calculations. In this gauge, we set $\xi^{0}=t,\,\xi^{1}=x$, and consider $r$ as a function of $x$ now. Accordingly, we can express the total action as follows:
\begin{align}
    S=gt&\Big(\int_{-\frac{L}{2}}^{0} w(r)\sqrt{g_{1}(r)+g_{2}(r)(\partial_{x}r)^2} dx+\int_{0}^{\frac{L}{2}} w(r)\sqrt{g_{1}(r)+g_{2}(r)(\partial_{x}r)^2}dx\notag\\
    &+3k\frac{e^{-2\mathbf{s}r^2}}{r} \sqrt{g_{1}(r)}+n\frac{e^{\frac{\mathbf{s}r^2}{2}}}{r} \sqrt{g_{1}(r)}\Big).
\end{align}
And now the boundary condition of $r(x)$ is
\begin{equation}
    r\Big(\pm\,\frac{L}{2}\Big)=0,\,r(0)=r_{v},\,\,
    \begin{cases}
    (\partial_{x}r)^2=\tan^2(\alpha),&r=r_{v}\\
      (\partial_{x}r)^2=0,&r=r_{0}.
    \end{cases}
\end{equation}

The generalized force balance equation at $r_{v}$ is the same as for intermediate $L$. We can use this equation to relate $r_{v}$ to $\alpha$. Additionally, we need to find the functional relationship between $r_{0}$ and $r_{v}$ due to inflection points. To accomplish this, we can employ the Euler-Lagrange equation and incorporate the action of the string to derive the first integral.

\begin{equation}
    \mathcal{H}=\frac{w(r)g_{1}(r)}{\sqrt{g_{1}(r)+g_{2}(r)(\partial_{x}r)^2}},
\end{equation}
$\mathcal{H}$ is a constant. We bring the boundary conditions to the first integral.
\begin{align}
    \mathcal{H}|_{r=r_{v}}&=\frac{w(r_{v})g_{1}(r_{v})}{\sqrt{g_{1}(r_{v})+g_{2}(r_{v})\tan^2(\alpha)}},\\
    \mathcal{H}|_{r=r_{0}}&=w(r_{0})\sqrt{g_{1}(r_{0})}.
\end{align}
So we obtain
\begin{gather}
    \partial_{x}r=\sqrt{\frac{w(r)^{2}g_{1}(r)^{2}-g_{1}(r)w(r_{0})^{2}g_{1}(r_{0})}{w(r_{0})^{2}g_{1}(r_{0})g_{2}(r)}}.
\end{gather}
Since the correspondence between $\alpha$ and $r_{v}$ has been established in the generalized force equilibrium equation, we can determine the correspondence between $r_{0}$ and $r_{v}$ by using the following relationship:
\begin{equation}
    w(r_{0})\sqrt{g_{1}(r_{0})}=\frac{w(r_{v})g_{1}(r_{v})}{\sqrt{g_{1}(r_{v})+g_{2}(r_{v})\tan^2(\alpha)}}.\label{41}
\end{equation}
Since $\partial_{r}x=\frac{1}{\partial_{x}r}$. To simplify the calculations, we can obtain the energy after performing the integral equivalent transformation and renormalization as follows:
\begin{align}
    E_{QQq}=g&\Big(2\int_{0}^{r_{v}} \big(w(r)\sqrt{g_{1}(r)(\partial_{r}x)^2+g_{2}(r)}-\frac{1}{r^{2}}\big)dr+2\int_{r_{v}}^{r_{0}} w(r)\sqrt{g_{1}(r)(\partial_{r}x)^2+g_{2}(r)}dr\notag\\
    &-\frac{2}{r_{v}}
    +3k\frac{e^{-2\mathbf{s}r_{v}^2}}{r} \sqrt{g_{1}(r_{v})}+n\frac{e^{\frac{\mathbf{s}r_{v}^2}{2}}}{r_{v}} \sqrt{g_{1}(r_{v})}\Big)+2c.\label{42}
\end{align}
Referring to Fig.~\ref{fig3}, the separation distance of the heavy-quark pair $L$ should be:
\begin{equation}
    L=2\Big(\int_{0}^{r_{v}}\frac{\partial x}{\partial r} dr+\int_{r_{v}}^{r_{0}}\frac{\partial x}{\partial r} dr\Big).\label{43}
\end{equation}
Combining Eqs. (\ref{42}) and (\ref{43}), we can obtain the potential energy at large $L$.

\section{Numerical results}\label{sec:03}

At low temperatures and low rapidities, the $\mathrm{QQq}$ is confined, there is an imaginary wall $r_{w}$ \cite{Cao:2022csq, Cao:2022mep, Yang:2015aia, Andreev:2006nw, Colangelo:2010pe}, but as the temperature or rapidity increases, the $\mathrm{QQq}$ will also become deconfined. When the $\mathrm{QQq}$ is confined, the separation distance $L$ and the potential energy $E$ will increase with $r_{v}$, and at a certain point $L$ will increase sharply until infinity. But in reality, the separation distance $L$ does not continue to increase infinitely. When the potential energy of the $\mathrm{QQq}$ reaches a certain critical point, the string connecting the quarks breaks, called string breaking. This leads to the excitation of light and anti-light quarks in the vacuum. When the $\mathrm{QQq}$ is deconfined, the imaginary wall disappears. As $r_{v}$ increases, the separation distance reaches a maximum, and simultaneously the potential energy reaches a maximum at this point, which means the $\mathrm{QQq}$ screening at this point.

This section examines the properties of $\mathrm{QQq}$ at finite rapidity from different research perspectives. It is divided into three parts: firstly, analyzing the influence of rapidity on the string breaking of the confined state; secondly, investigating the effect of screening when the $\mathrm{QQq}$ is deconfined; and finally, conducting a comprehensive comparison with $\mathrm{Q\Bar{Q}}$. Additionally, we construct a state diagram of $\mathrm{QQq}$ and $\mathrm{Q\Bar{Q}}$ in the $T-\eta$ plane to graphically represent their behavior.

\begin{figure}
    \centering
    \includegraphics[width=8.5cm]{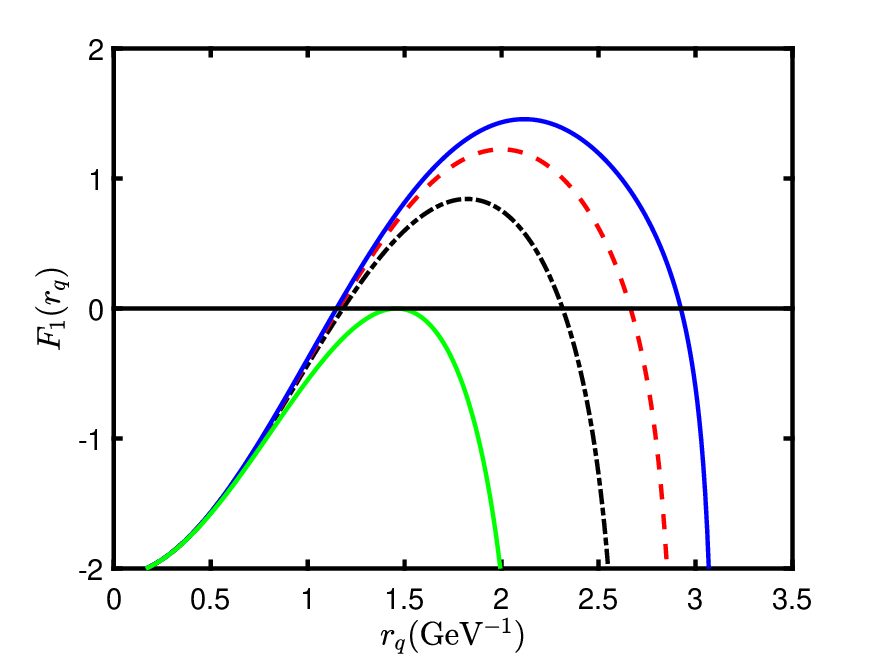}
    \caption{\label{fig4} $F_{1}$ as a function of $r_{q}$ for various rapidities at $T=0.1\,\mathrm{GeV}$. The blue line represents $\eta=0.3$; the red dashed line represents $\eta=0.6$; the black dot-dashed line represents $\eta=0.9$; the green line represents $\eta=1.3865$. Only the green line has one zero point, and the other three lines have two zero points.}
\end{figure}

From Eq. (\ref{23}) we can solve for the light quark position corresponding to $(T,\eta)$. There are two solutions to this equation within the valid interval, and only the smaller solution meets our requirements. We consider this equation as a function of $r_{q}$ with different $T$ and $\eta$ shown in Fig.~\ref{fig4}. We solve for the $r_{q}$ in these cases at $T=0.1\,\mathrm{GeV}$, and obtain the results respectively: $\eta=0.3,\,r_{q}=1.1487\,\mathrm{GeV^{-1}};\,\eta=0.6,\,r_{q}=1.1587\,\mathrm{GeV^{-1}};\,\eta=0.9,\,r_{q}=1.1823\,\mathrm{GeV^{-1}};\,\eta=1.3865,\,r_{q}=1.4582\,\mathrm{GeV^{-1}}$.

In Fig.~\ref{fig4}, it can be observed that at a constant temperature, the function $F_{1}(r_{q})$ decreases as the rapidity increases until it no longer has any zero values. For each temperature value, there is a corresponding rapidity that results in the function having exactly one solution. We can obtain solutions by solving the following equations:
\begin{equation}
    \begin{cases}
    F_{1}(r_{q},T,\eta )&=0,\\
    \frac{\partial F_{1}(r_{q},T,\eta)}{\partial r_{q}}&=0.
    \end{cases}.\label{44}
\end{equation}

\begin{figure}
    \centering
    \includegraphics[width=8.5cm]{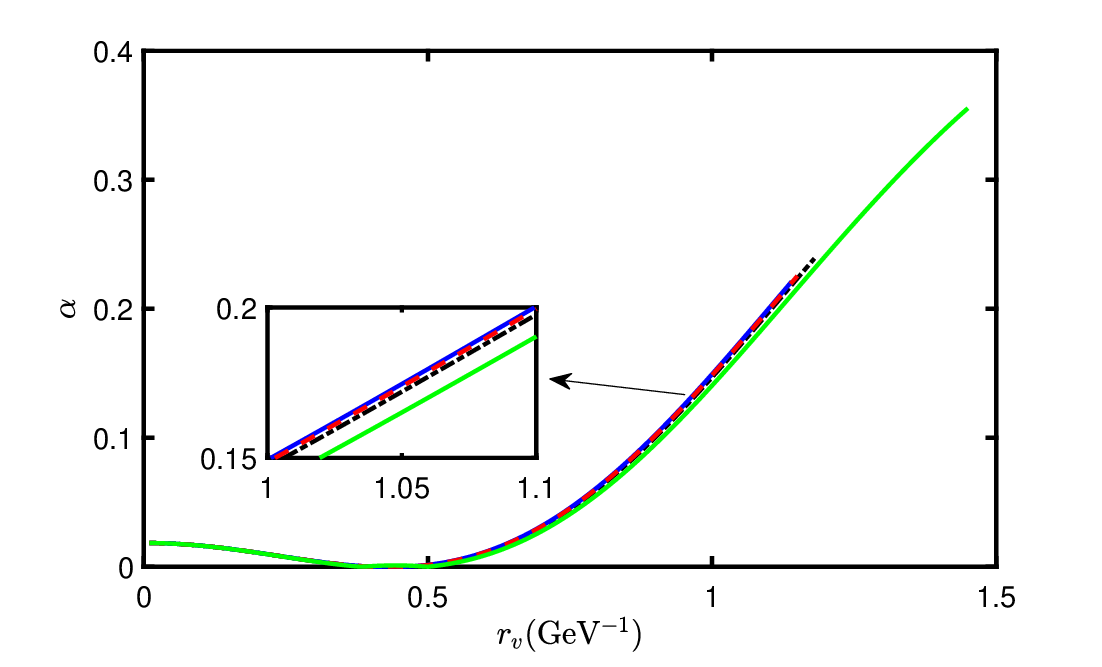}
    \caption{\label{fig6} The relation between the position $r_{v}$ of the baryon vertex and the angle $\alpha$, which is between the string and the horizon, at $T=0.1\,\mathrm{GeV}$ and various rapidities. Among them, the blue line represents $\eta=0.3$; the red dashed line represents $\eta=0.6$; the black dot-dashed line represents $\eta=0.9$; the green line represents $\eta=1.3865$.}
\end{figure}
We calculate the angle at different rapidities using Eq. (\ref{24}), as a function of $r_{v}$ at a temperature of $T=0.1\,\mathrm{GeV}$. The resulting plot is shown in Fig.~\ref{fig6}.  It is evident that regardless of the parameters, the angles are always greater than zero, and as the rapidity increases, the angle decreases. The transverse axes of the angles all start at zero and end at their respective light quark positions $r_{q}$.  When the baryon vertex coincides with the light quark, denoted by $r_{v}=r_{q}$,  it indicates that the configuration has entered the second stage, which is the intermediate $L$. Subsequently, the light quark and the baryon vertex rise together. The stage where $r_{v} < r_{q}$ is referred to as the first stage, known as the small $L$.

\begin{figure}
    \centering
    \includegraphics[width=8.5cm]{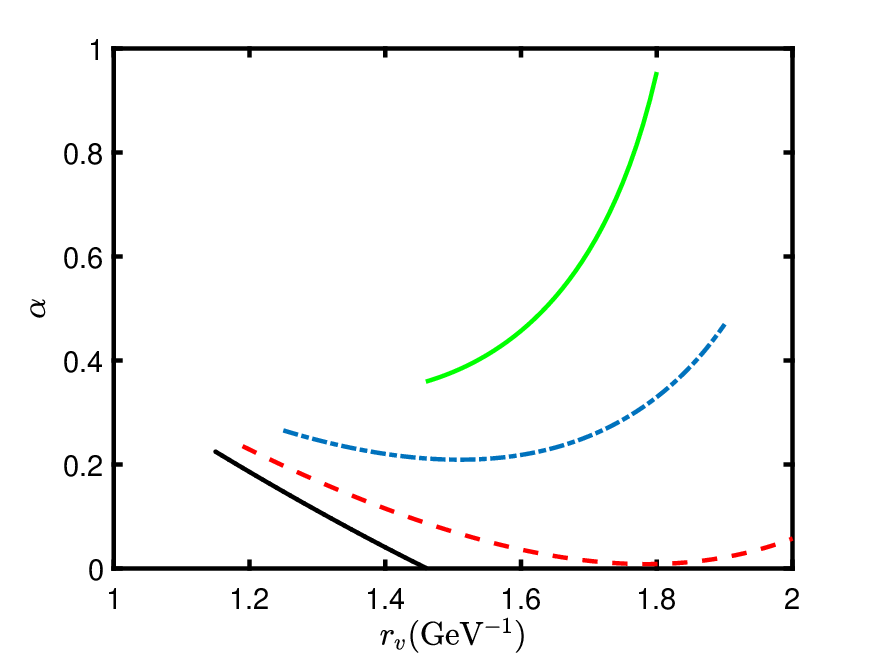}
    \caption{\label{fig7} The relation between the position $r_{v}$ of the baryon vertex and the angle $\alpha$, which is between the string and the horizontal, at $T=0.1\,\mathrm{GeV}$ and various rapidities. Among them, the black line represents $\eta=0.3$; the red dashed line represents $\eta=0.9$; the blue dot-dashed line represents $\eta=1.2$; the green line represents $\eta=1.3865$.}
\end{figure}

At the intermediate $L$, we can solve Eq. (\ref{30}) to obtain the relation between the angles $\alpha$ and $r_{v}$. Then we draw this plot at different rapidities, as shown in Fig.~\ref{fig7}. It can be seen that when the rapidity is small, the angle decreases with $r_{v}$ linearly, and when it decreases to less than $0$, we consider that $\mathrm{QQq}$ transitions from the second stage to the third stage, which is large $L$. In addition, as the rapidity increases, the angular plot of the second stage changes greatly. The curve of $\alpha$ always moves upwards, no longer monotonically decreasing, and it gradually bends upwards from a straight line to a curve, ultimately becoming monotonically increasing. From the results, when the rapidity is large, the third stage configuration of the model will not be possible.

\begin{figure}
    \centering
    \includegraphics[width=8.5cm]{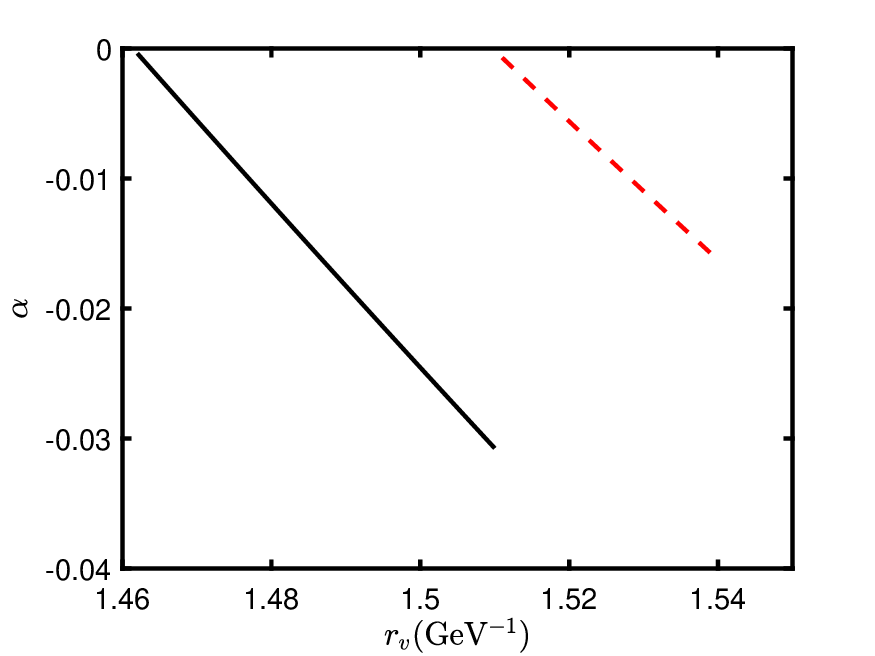}
    \caption{\label{fig8} The relation between the position $r_{v}$ of the baryon vertex and the angle $\alpha$, which is between the string and the horizontal, at $T=0.1\,\mathrm{GeV}$ and various rapidities. Among them, the black line represents $\eta=0.3$; the red dashed line represents $\eta=0.6$. }
\end{figure}

\begin{figure}
    \centering
    \includegraphics[width=8.5cm]{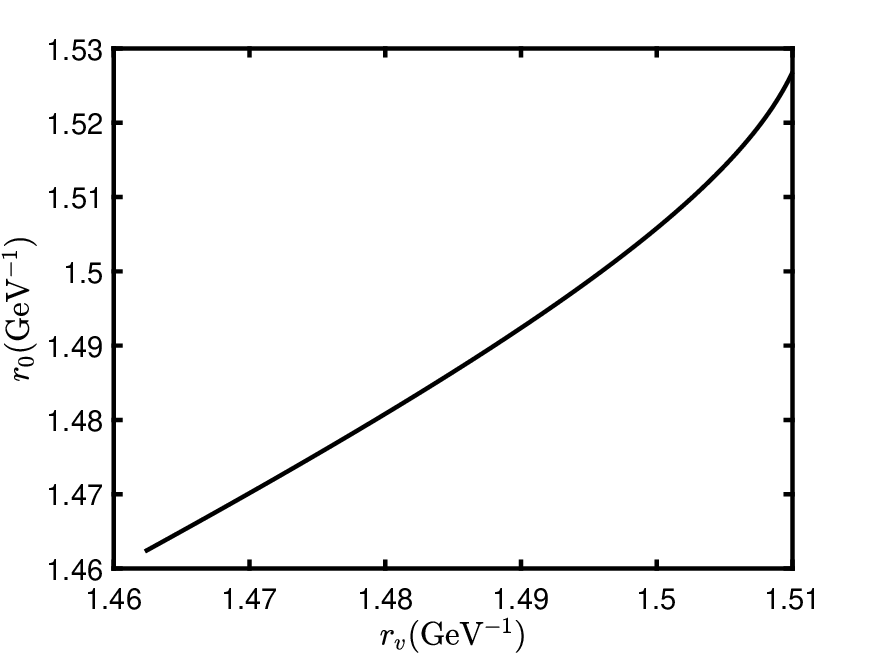}
    \caption{\label{fig9} The relation between the smooth turning point $r_{0}$ on the string and the position $r_{v}$ of the baryon vertex at $T=0.1\,\mathrm{GeV}$, $\eta=0.3$.}
\end{figure}

Fig.~\ref{fig8} shows the relation between the angles $\alpha$ and the $r_{v}$ in the third stage. This result is obtained through Eq. (\ref{30}). Their transverse axes all start at the end point of their respective second stages, and the longitudinal axes start at $0$. It can be seen that as the rapidity increases, the range of configuration in the third stage continues to decrease until it no longer exists.  This dovetails with our analysis in the second phase. Since the string configuration features smooth turning points during the third stage, we obtain the relation between $r_{v}$ and $r_{0}$ using Eq. (\ref{41}), as shown in Fig.~\ref{fig9}. From the calculated results between $\alpha$ and $r_{v}$ in each stage, it can be observed that $\alpha$ is smooth between the two stages.

\begin{figure}
    \centering
    \includegraphics[width=8.5cm]{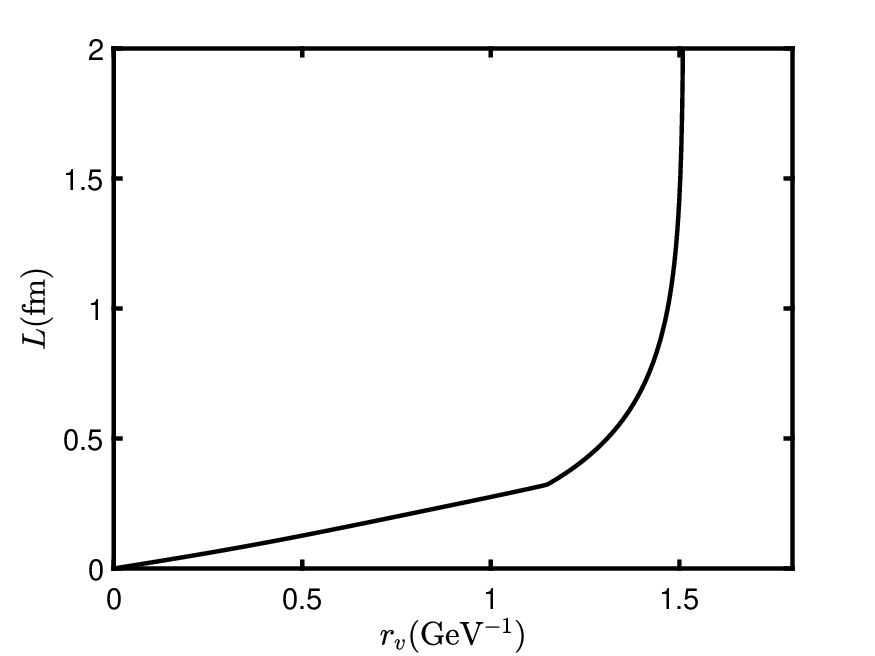}
    \caption{\label{fig10} The separation distance $L$ between pairs of heavy quarks as a function of the position $r_{v}$ of baryon vertex at $T=0.1\,\mathrm{GeV}$, $\eta=0.3$.}
\end{figure}

\begin{figure}
    \centering
    \includegraphics[width=8.5cm]{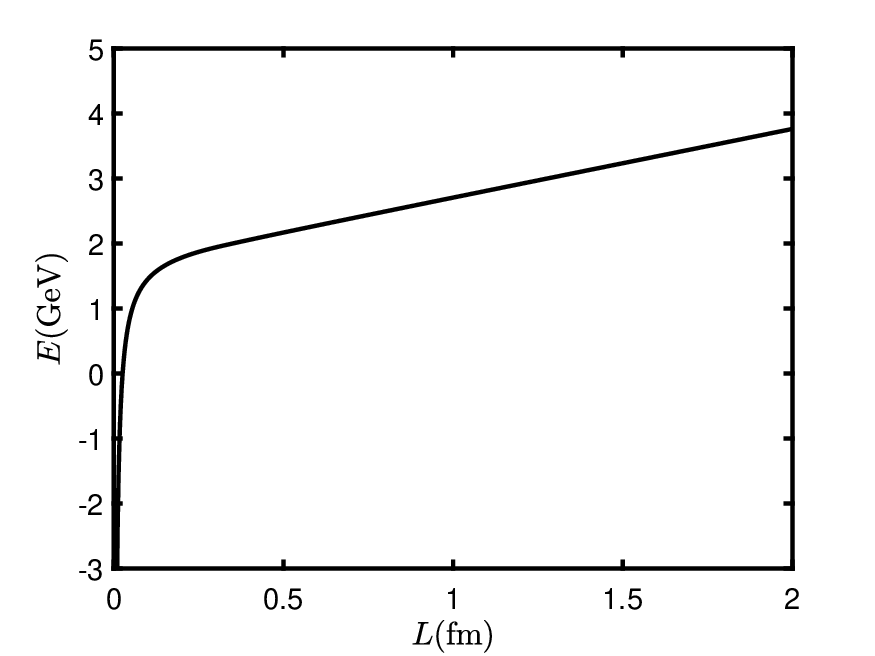}
    \caption{\label{fig11} The relation between $\mathrm{QQq}$ potential energy $E$ and heavy-quark pair separation distance $L$ at $T=0.1\,\mathrm{GeV}$, $\eta=0.3$.}
\end{figure}

In Fig.~\ref{fig10}, we present the results of $L$ as a function of $r_v$ based on Eqs. (\ref{25}) and (\ref{43}). At $T=0.1\,\mathrm{GeV}$, $\eta=0.3$, the QQq will be confined and the separation distance can tend to infinity. Subsequently, we use Eqs. (\ref{13}), (\ref{29}), and (\ref{42}) to obtain a plot of $E$ and $L$ as shown in Fig.~\ref{fig11}. In the confined state, the potential of $\mathrm{QQq}$ is also the Cornell potential with finite temperature and rapidity.

\subsection{Confined state} \label{A}

When the $\mathrm{QQq}$ is confined, string breaking occurs when potential energy $E$ reaches a certain value, exciting positive and negative light quarks in the vacuum. The pattern of string breaking is as follows:
\begin{equation}
    \mathrm{QQq} \longrightarrow \mathrm{Qqq}+\mathrm{Q\Bar{q}}.
\end{equation}

\begin{figure}
    \centering
    \includegraphics[width=8.5cm]{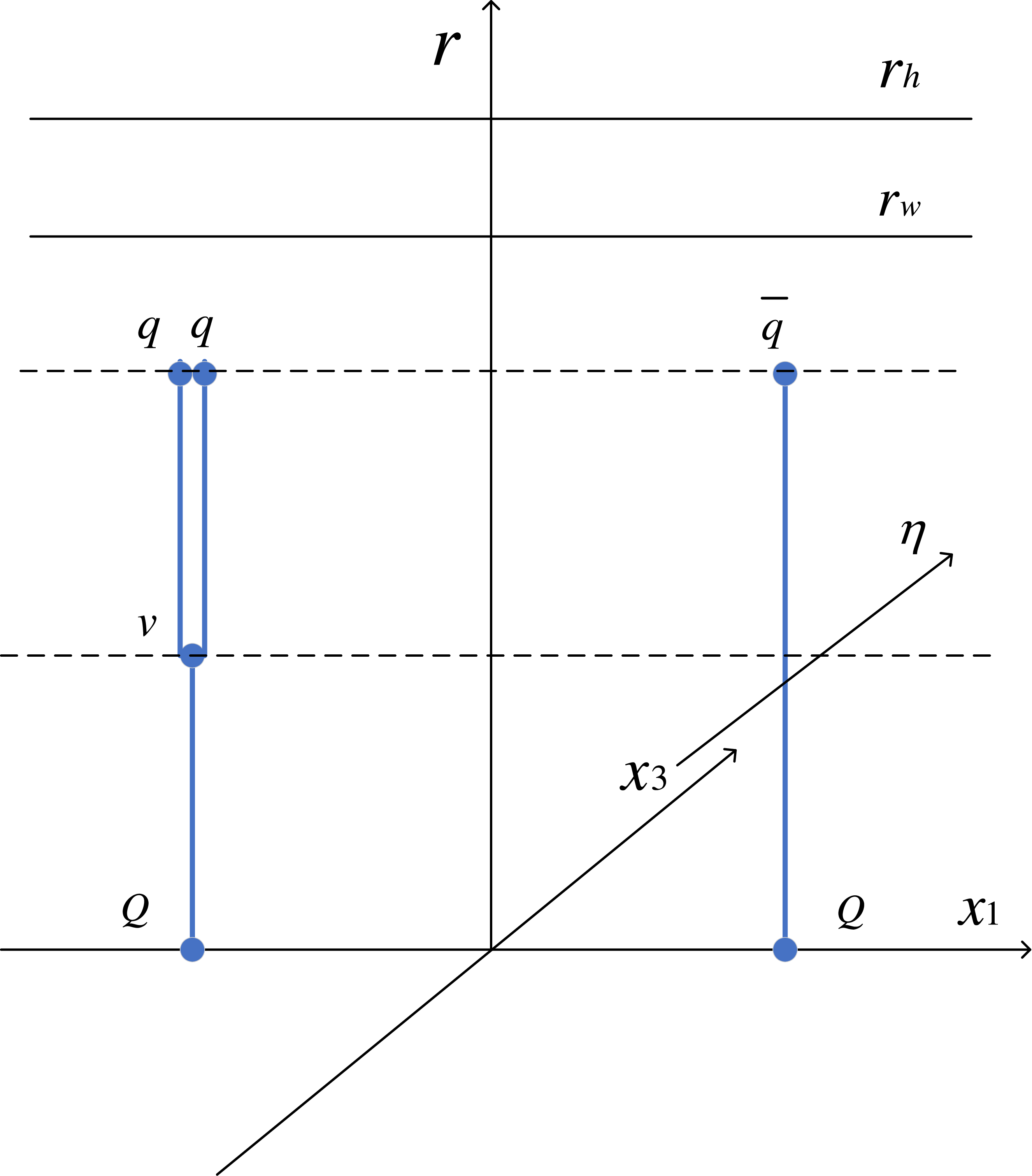}
    \caption{\label{fig12} Schematic diagram after string breaking. The rapidity $\eta $ is along the {\color{red} $x_3$-axis} direction. $r_{h}$ is the position of the black hole horizon. $r_{w}$ is the position of an imaginary wall when the $\mathrm{QQq}$ is confined.}
\end{figure}

The schematic diagram after the string breaking is Fig.~\ref{fig12}. It can be seen that the action of $\mathrm{Qqq}$ is provided by three strings, one baryon vertex, and two light quarks. The action of $\mathrm{Q\Bar{q}}$ consists of a string and a light quark. Therefore, the total action after the string breaking is:

\begin{align}
    S_{Qqq}&=\sum_{i=1}^{3} S_{NG}^{(i)}+S_{vert}+2S_{q}, \\
    S_{Q\Bar{q}}&=S_{NG}+S_{q}.
\end{align}
We choose the gauge as $\xi^{0}=t,\,\xi^{1}=r$, therefore, we obtain that the boundary condition of $x(r)$ is
\begin{equation}
    (\partial_{r}x)^2=0.
\end{equation}
So the total action can be written as
\begin{align}
    S=gt&\Big(\int_{0}^{r_{v}}w(r)\sqrt{g_{2}(r)}dr+2\int_{r_{v}}^{r_{q}}w(r)\sqrt{g_{2}(r)}dr+\int_{0}^{r_{q}}w(r)\sqrt{g_{2}(r)}dr\notag\\
    &+3k\frac{e^{-2\mathbf{s}r_{v}^2}}{r_{v}} \sqrt{g_{1}(r_{v})}+3n\frac{e^{\frac{\mathbf{s}r_{q}^2}{2}}}{r_{q}} \sqrt{g_{1}(r_{q})}\Big).
\end{align}

Simple analysis shows that the generalized force equilibrium equation at light quarks does not change after string breaking, that is, the positions of the three light quarks will be consistent with the one before breaking. However, the force configuration at the baryon vertex of $\mathrm{Qqq}$ changes, and we can denote it as follows:
\begin{equation}
    F_{2}(r_{v})=\frac{2e_{3}-e_{3}+f_{v}}{gw(r_{v})}=\frac{e_{3}+f_{v}}{gw(r_{v})}=0,\label{50}
\end{equation}
where $e_{3}$ and $f_{v}$ coincide with Eqs. (\ref{19}) and (\ref{22}). There is only one unknown quantity $r_{v}$ in the above equation, so we can solve for its value. This equation usually solves for two values within the range where $r_{v}$ has physical significance, and we take the smaller one here. We solve for the $r_{v}$ in these cases at $T=0.1\,\mathrm{GeV}$,  respectively: $\eta=0.3,\,r_{v}=0.4088\,\mathrm{GeV^{-1}};\,\eta=0.6,\,r_{v}=0.4049\,\mathrm{GeV^{-1}};\,\eta=0.9,\,r_{v}=0.3987\,\mathrm{GeV^{-1}}$.

Then we consider the renormalize to find the potential energy after the string breaking:
\begin{align}
    E_{break}=g&\Big(\int_{0}^{r_{v}}\big(w(r)\sqrt{g_{2}(r)}-\frac{1}{r^{2}}\big)dr+2\int_{r_{v}}^{r_{q}}w(r)\sqrt{g_{2}(r)}dr+\int_{0}^{r_{q}}\big(w(r)\sqrt{g_{2}(r)}-\frac{1}{r^2}\big)dr\notag\\
    &-\frac{1}{r_{v}}-\frac{1}{r_{q}}
    +3k\frac{e^{-2\mathbf{s}r_{v}^2}}{r_{v}} \sqrt{g_{1}(r_{v})}+3n\frac{e^{\frac{\mathbf{s}r_{q}^2}{2}}}{r_{q}} \sqrt{g_{1}(r_{q})}\Big)+2c.
\end{align}
We calculate the potential energy at $T=0.1\,\mathrm{GeV}$ after string breaking and obtain: $\eta=0.3,\, E_{break}=3.0040\,\mathrm{GeV}; \,\eta=0.6,\, E_{break}=2.9983\,\mathrm{GeV}; \,\eta=0.9,\, E_{break}=2.9857\,\mathrm{GeV}$.

\begin{figure}
    \centering
    \includegraphics[width=8.5cm]{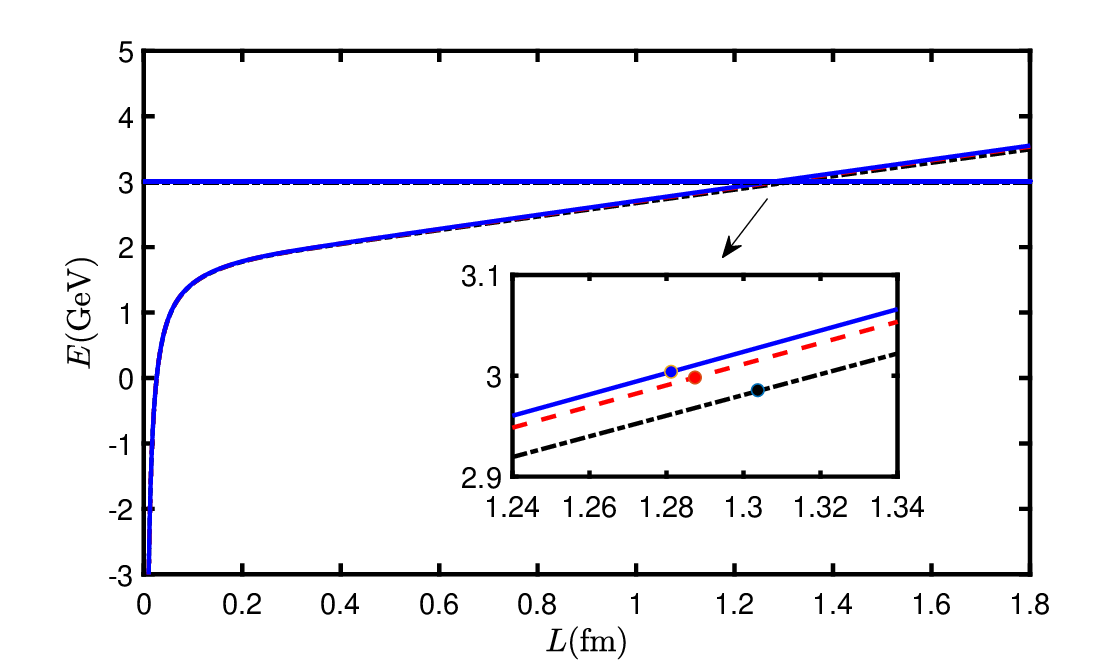}
    \caption{\label{fig13} The curves in the figure are the potential energy of $\mathrm{QQq}$ at $T=0.1\,\mathrm{GeV}$, and the straight lines are the potential energy at $T=0.1\,\mathrm{GeV}$ after string breaking. The black dot-dashed line represents $\eta=0.9$. The red dashed line represents $\eta=0.6$. The blue solid line represents $\eta=0.3$. }
\end{figure}

The potential energies of $\mathrm{QQq}$ is shown in Fig.~\ref{fig13}. From Fig.~\ref{fig13}, it can be observed that a higher rapidity leads to a lower potential energy of $\mathrm{QQq}$.
The calculated data is as follows: $\eta=0.3,\,L_{break}=1.2812\,\mathrm{fm};\,\eta=0.6,\,L_{break}=1.2874\,\mathrm{fm};\,\eta=0.9,\,L_{break}=1.3037\,\mathrm{fm}$. Based on these results, we conclude that the string-breaking distance is enhanced and the potential energy is decreased in the presence of increased rapidity.

\subsection{Deconfined state} \label{B}

\begin{figure}
    \centering
    \includegraphics[width=8.5cm]{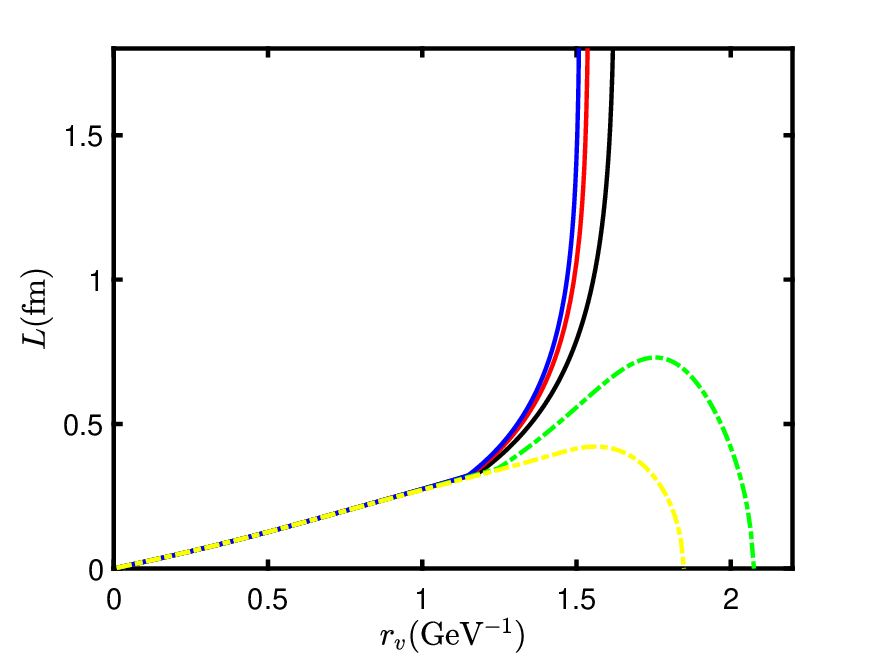}
    \caption{\label{fig14} The separation distance $L$ between pairs of heavy quarks as a function of the position $r_{v}$ of the baryon vertex at $T=0.1\,\mathrm{GeV}$. The solid line represents in confined state, while the dashed line represents in deconfined state. The blue line represents $\eta=0.3$; the red line represents $\eta=0.6$; the black line represents $\eta=0.9$; the green line represents $\eta=1.2$, and the yellow line represents $\eta=1.3865$. }
\end{figure}

The confined or deconfined state of the $\mathrm{QQq}$ can be determined by analyzing the plots of $L-r_{v}$ and the effective string tension. It has been observed that the results obtained from both methods are consistent \cite{Andreev:2006nw}. We calculate the plots of $L-r_{v}$ for different rapidities at the same temperature, as shown in Fig.~\ref{fig14}. As the separation distance $L$ of the $\mathrm{QQq}$ increases, it approaches infinity. On the other hand, when it is deconfined, the $L-r_{v}$ plot has a maximum value. $\mathrm{QQq}$ will become free quarks above the maximum value.

Fig.~\ref{fig14} illustrates the transition of the $\mathrm{QQq}$ from a confined to a deconfined state at a critical rapidity and temperature. When the $\mathrm{QQq}$ is deconfined, it is observed that the greater the rapidity, the smaller the screening distance and $r_{v}$. This implies that as the rapidity increases, the $\mathrm{QQq}$ becomes easier to screen. When $T=0.1\,\mathrm{GeV},\, \eta=1.3865$ is the maximum rapidity that the $\mathrm{QQq}$ model can allow, and its corresponding screening distance is $L_{melt}=0.4218\,\mathrm{fm}$, then $r_{v}=1.5682\,\mathrm{GeV^{-1}}$.

\begin{figure}
    \centering
    \includegraphics[width=8.5cm]{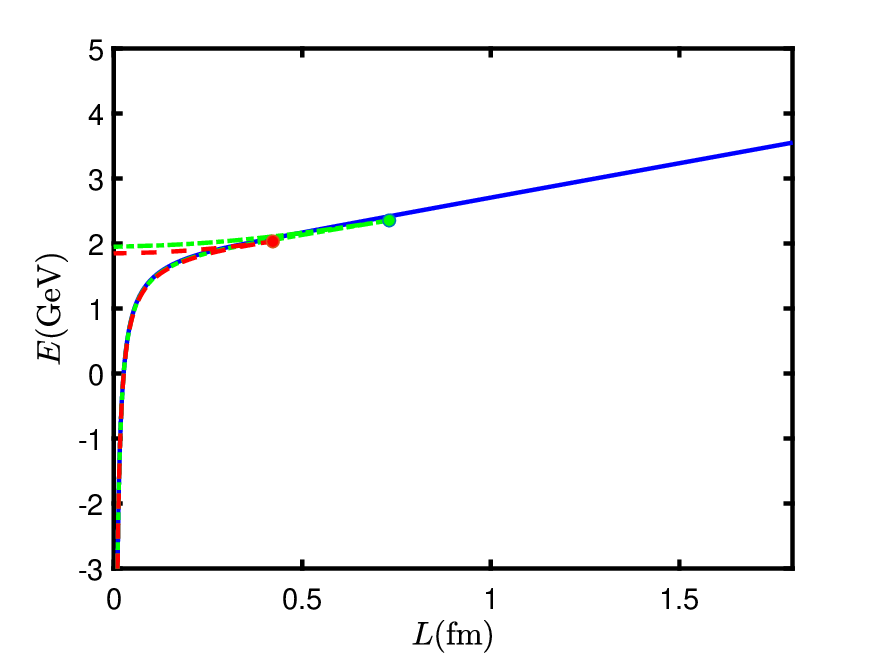}
    \caption{\label{fig15} The relation between the potential energy $E$ and the separation distance $L$ at $T=0.1\,\mathrm{GeV}$. The solid line represents in confined state, while the dashed line represents in deconfined state. The solid blue line represents $\eta =0.3$; the green dot-dashed line represents $\eta =1.2$; the red dashed line represents $\eta =1.3865$. }
\end{figure}

We also calculate the plots of $E$ and $L$ as shown in Fig.~\ref{fig15}. From the figure, it can be seen that the potential energy of the confined $\mathrm{QQq}$ steadily increases with $L$ until string breaking happens. There is a maximum value for the potential energy of the deconfined state. When $E$ reaches the maximum value, $L$ also reaches the maximum value, which is the apex of the $L$ in Fig.~\ref{fig14}. As the rapidity increases, the potential energy of $\mathrm{QQq}$ decreases, and the screening distance becomes smaller. Fig.~\ref{fig15} also illustrates that, regardless of whether it is in the confined or deconfined state, as the rapidity increases at a fixed temperature, the potential energy becomes smaller. The calculation results about the screening of the $\mathrm{QQq}$ in the deconfined state are as follows: at $\eta =1.2$, $L_{screen}=0.7304\,\mathrm{fm},\, E_{screen}=2.3547\,\mathrm{GeV}$; at $\eta =1.3865$, $L_{screen}=0.4219\,\mathrm{fm},\, E_{screen}=2.0317\,\mathrm{GeV}$.

\begin{figure}
    \centering
    \includegraphics[width=8.5cm]{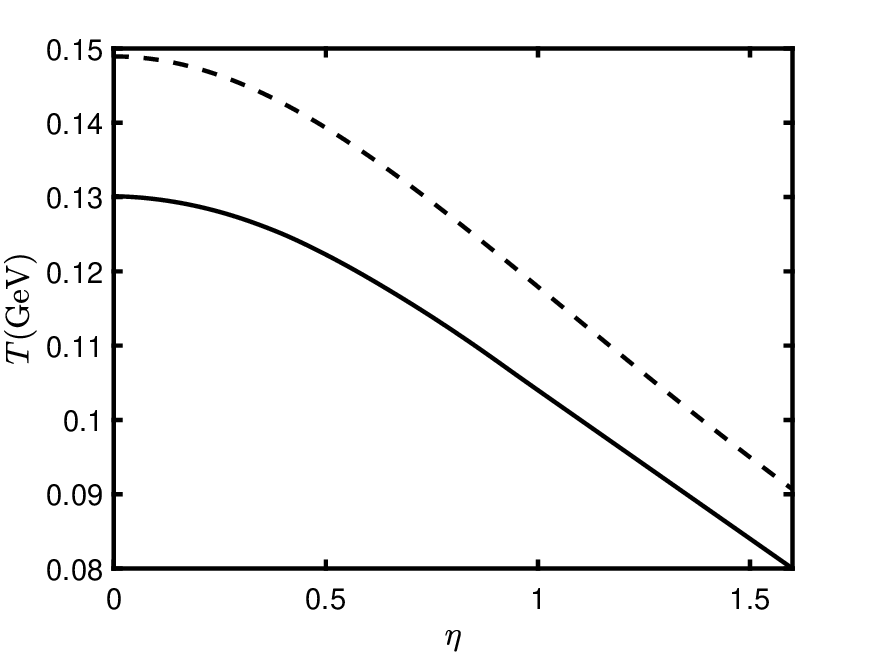}
    \caption{\label{fig16} The dashed line consists of the maximum $(\eta,\, T)$ points determined by the Eqs. (\ref{44}), and the solid line consists of the critical points that distinguish the confined state from the deconfined state. }
\end{figure}

We will briefly introduce the way used to determine the confined and deconfined state. If the maximum value of the screening distance is smaller than the distance of string breaking, it indicates that the $\mathrm{QQq}$ is in the deconfined state; otherwise, it is in the confined state. Using this method, we perform calculations to determine the critical points, and these results are plotted in Fig.~\ref{fig16}.

In Fig.~\ref{fig16}, $\mathrm{QQq}$ cannot exist in a medium with temperatures and rapidities above the dashed line. The $\mathrm{QQq}$ is in the deconfined state between the dashed and solid lines. In this state, $\mathrm{QQq}$ become free quarks at long distances due to screening. When $\mathrm{QQq}$ is in the deconfined state, with the increase of the rapidity at a fixed temperature, the screening distance decreases. Below the solid line, the $\mathrm{QQq}$ is in the confined state. In this state, when the separation distance reaches the string-breaking distance and string breaking occurs, it transitions into a new configuration.

\subsection{Comparison with $\mathrm{Q\Bar{Q}}$
configuration} \label{C}

Research on $\mathrm{Q\Bar{Q}}$ through holographic models is mature enough \cite{Andreev:2021vjr, Andreev:2006eh, Ewerz:2016zsx, He:2010bx, Colangelo:2010pe, Li:2011hp, Fadafan:2011gm, Fadafan:2012qy, Cai:2012xh}. We mainly study the difference between $\mathrm{Q\Bar{Q}}$ and $\mathrm{QQq}$ under the influence of rapidities.

First of all, we can write the metric after adding the rapidity effect of $\mathrm{Q\Bar{Q}}$ through the Lorentz transformation:
\begin{align}
    ds^{2}=w(r)&\Big(-g_{1}(r)dt^{2}
 -2\sinh(\eta)\cosh(\eta)\big(1-\frac{g_{1}(r)}{g_{2}(r)}\big)dx_{3}dt\notag\\ &+g_{3}(r)dx_{3}^{2}+dx_{1}^{2}+dx_{2}^{2}+\frac{g_{2}(r)}{g_{1}(r)}dr^{2}\Big),
\end{align}
where $w(r),\,g_{1}(r),\,g_{2}(r),$ and $\,g_{3}(r)$ are consistent with the previously derived Eqs. (\ref{2}) and (\ref{5}).

We select the static gauge $\xi^{0}=t$, $\xi^{1}=x$, and the action of $\mathrm{Q\Bar{Q}}$ can be written as
\begin{gather}
    S=gt\Big(\int_{-\frac{L}{2}}^{0} w(r)\sqrt{g_{1}(r)+g_{2}(r)(\partial_{x}r)^2} dx+\int_{0}^{\frac{L}{2}} w(r)\sqrt{g_{1}(r)+g_{2}(r)(\partial_{x}r)^2}\Big)dx.
\end{gather}
And now the boundary condition of $r(x)$ is
\begin{equation}
    r\Big(\pm\,\frac{L}{2}\Big)=0,\,r(0)=r_{0},\,(\partial_{x}r|_{r=r_{0}})^2=0.
\end{equation}
The string configuration of $\mathrm{Q\Bar{Q}}$ is U-shaped, where $r_{0}$ is the smooth turning point of the U-shaped string.
With the Euler Lagrange equation, we can obtain
\begin{equation}
    \partial_{x}r=\sqrt{\frac{w(r)^{2}g_{1}(r)^{2}-g_{1}(r)w(r_{0})^{2}g_{1}(r_{0})}{w(r_{0})^{2}g_{1}(r_{0})g_{2}(r)}}.
\end{equation}
By renormalizing, we can obtain the potential energy of $\mathrm{Q\Bar{Q}}$,
\begin{gather}
    E_{Q\Bar{Q}}=g\Big(2\int_{0}^{r_{0}} \big(w(r)\sqrt{g_{1}(r)(\partial_{r}x)^2+g_{2}(r)}-\frac{1}{r^{2}}\big)dr-\frac{2}{r_{0}}\Big)+2c.
\end{gather}
And,
\begin{equation}
    L=2\int_{0}^{r_{0}}\frac{\partial x}{\partial r}  dr,
\end{equation}
where $\partial_{r}x=\frac{\partial x}{\partial r}=\frac{1}{\partial_{x}r}$.

\begin{figure}
    \centering
    \includegraphics[width=8.5cm]{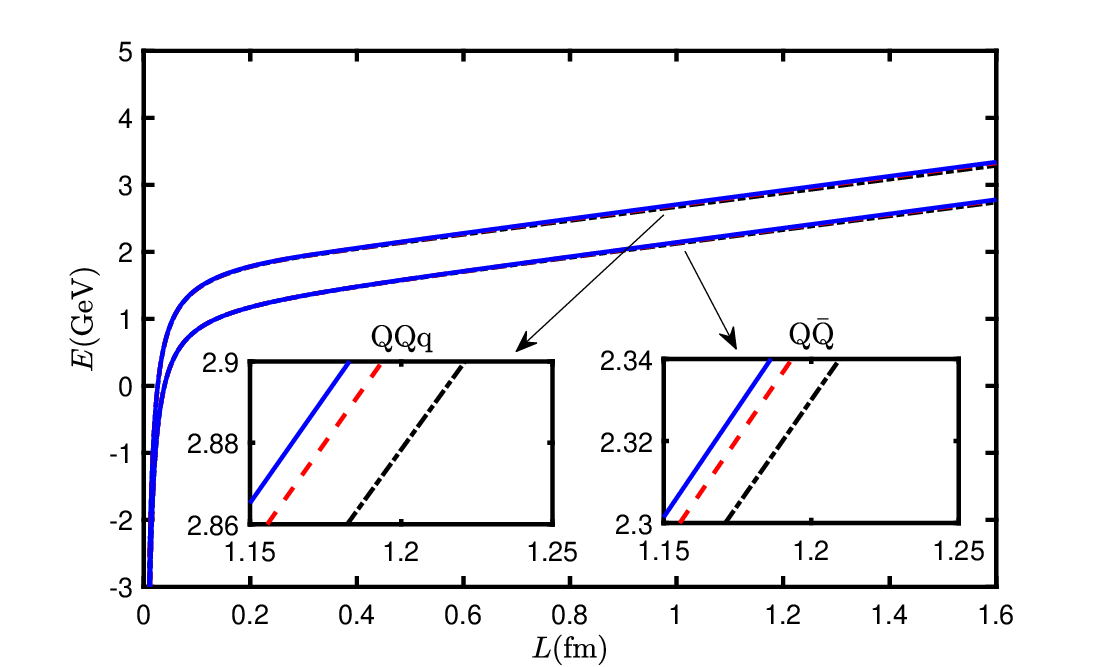}
    \caption{\label{fig18} The potential energies of $\mathrm{Q\Bar{Q}}$ and $\mathrm{QQq}$ are depicted at $T=0.1\,\mathrm{GeV}$. In the graph, the black dot-dashed line represents $\eta=0.9$; the red dashed line represents $\eta=0.6$; and the blue solid line represents $\eta=0.3$. }
\end{figure}

We can calculate the potential energy of $\mathrm{Q\Bar{Q}}$ at different rapidities at the same temperature and compare it with $\mathrm{QQq}$. The result is presented in Fig.~\ref{fig18}, indicating that the potential energy of $\mathrm{QQq}$ is higher than that of $\mathrm{Q\Bar{Q}}$. Furthermore, it is observed that increasing rapidity consistently leads to a decrease in the potential energy of $\mathrm{Q\Bar{Q}}$ and $\mathrm{QQq}$. When $\mathrm{Q\bar{Q}}$ is in the confined state, the $\mathrm{Q\bar{Q}}$ string breaks when the separation distance becomes large enough. Let us consider the pattern of $\mathrm{Q\bar{Q}}$ string breaking as
\begin{equation}
    \mathrm{Q\Bar{Q}} \longrightarrow \mathrm{Q\Bar{q}}+\mathrm{\Bar{Q}q}.
\end{equation}
It is easy to know that
\begin{equation}
    E_{break}=2g\Big(\int_{0}^{r_{q}}\big(w(r)\sqrt{g_{2}(r)}-\frac{1}{r^2}\big)dr-\frac{1}{r_{q}}+n\frac{e^{\frac{sr_{q}^2}{2}}}{r_{q}} \sqrt{g_{1}(r_{q})}\Big)+2c.
\end{equation}

\begin{figure}
    \centering
    \includegraphics[width=8.5cm]{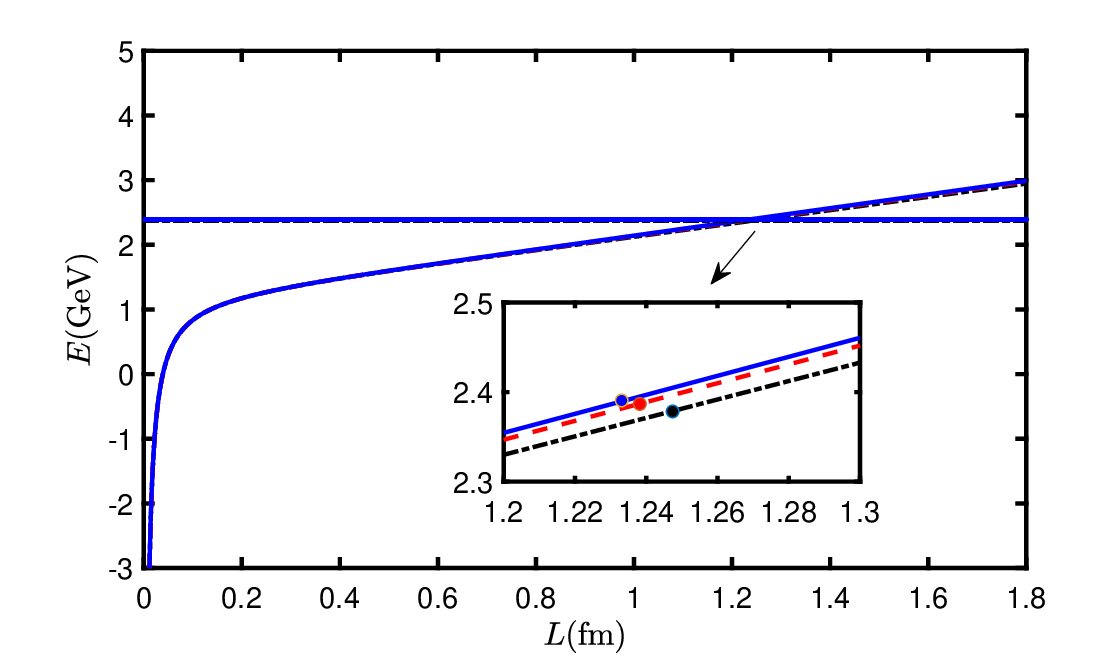}
    \caption{\label{fig19} The curves in the figure are the potential energy of $\mathrm{Q\Bar{Q}}$ at $T=0.1\,\mathrm{GeV}$, and the straight lines are the potential energy at $T=0.1\,\mathrm{GeV}$ after string breaking. Where the black dot-dashed line represents $\eta=0.9$; the red dashed line represents $\eta=0.6$; and the blue solid line represents $\eta=0.3$. }
\end{figure}

Based on this, we can draw Fig.~\ref{fig19}. As shown in the figure, it is evident that increasing rapidity leads to a greater separation distance at the point of string breaking for both $\mathrm{Q\Bar{Q}}$ and $\mathrm{QQq}$. The calculated data are $\eta=0.3,\, E_{break}=2.3906\,\mathrm{GeV},\, L_{break}=1.2331\,\mathrm{fm}; \,\eta=0.6,\, E_{break}=2.3868\,\mathrm{GeV},\, L_{break}=1.2382\,\mathrm{fm}; \,\eta=0.9,\, E_{break}=2.3784\,\mathrm{GeV},\, L_{break}=1.2474\,\mathrm{fm}$. It can be observed that the breaking distance of $\mathrm{Q\Bar{Q}}$ is slightly less than $\mathrm{QQq}$, but the difference between the two is not significant.

\begin{figure}
    \centering
    \includegraphics[width=8.5cm]{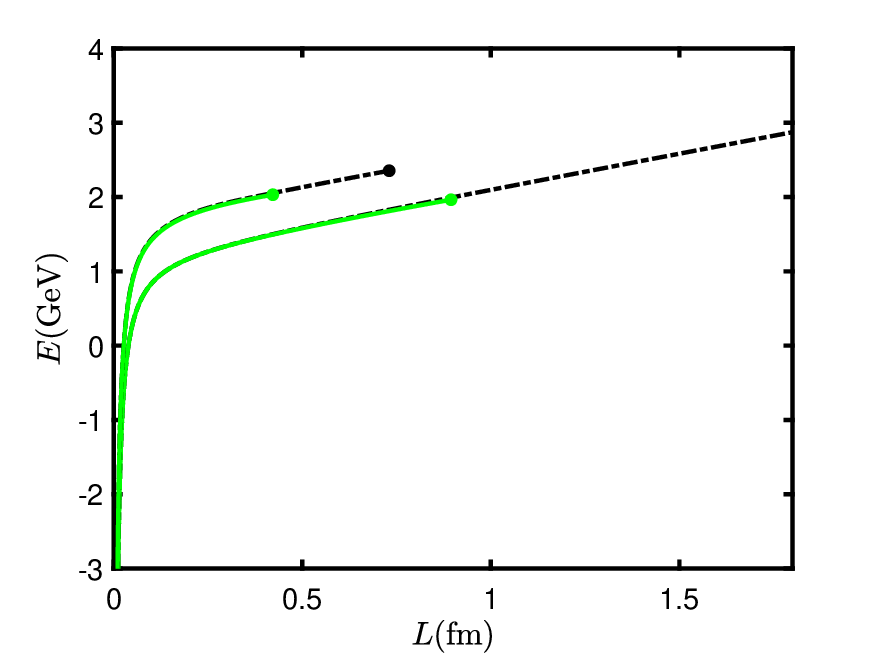}
    \caption{\label{fig20} The higher curves represent $\mathrm{QQq}$, and the lower curves represent $\mathrm{Q\Bar{Q}}$. The black line represents $\eta=1.2,\,T=0.1\,\mathrm{GeV}$, and the green line represents $\eta=1.3865,\,T=0.1\,\mathrm{GeV}$. }
\end{figure}

When the rapidity continues to increase, $\mathrm{Q\Bar{Q}}$ will also change from the confined state to the deconfined state. For this purpose, we calculated the results in Fig.~\ref{fig20}. When we calculated the potential of $\mathrm{QQq}$ at $\eta =1.3865$ and $T = 0.1GeV$, the results show $L_{melt}=0.4219\,\mathrm{fm}$ and $E_{melt}=2.0317\,\mathrm{GeV}$. When $\mathrm{Q\Bar{Q}}$ is at $\eta =1.3865$ and $T = 0.1GeV$, we get $L_{melt}=0.8944\,\mathrm{fm}$ and $E_{melt}=1.9640\,\mathrm{GeV}$. It can be found that the difference in screening distances is significant. We infer that $\mathrm{Q\Bar{Q}}$ is more stable than $\mathrm{QQq}$ at the same temperature and rapidity.

\begin{figure}
    \centering
    \includegraphics[width=8.5cm]{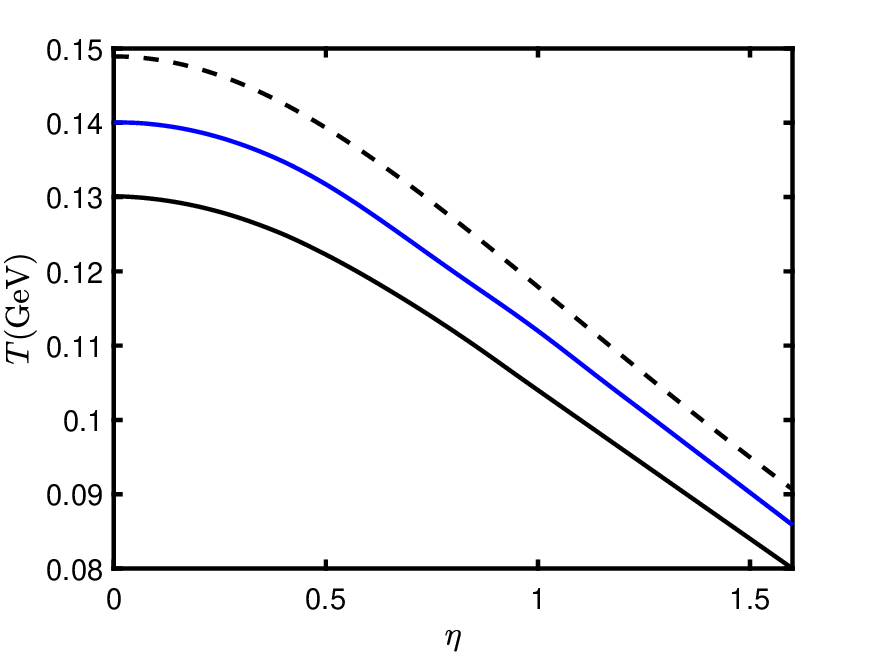}
    \caption{\label{fig21} The solid black line is to distinguish the confined state and the deconfined state of $\mathrm{QQq}$; the dashed line is the maximum conditions allowed by $\mathrm{QQq}$; the blue line is to distinguish the confined state and the deconfined state of $\mathrm{Q\Bar{Q}}$. }
\end{figure}

Fig.~\ref{fig21} shows  $\mathrm{QQq}$ is confined below the black line, which means the separation distance of heavy quarks can be infinite if we ignore the string breaking. $\mathrm{QQq}$ can still be found below the black dashed line even the $\mathrm{QQq}$ is in the deconfined state which means the $\mathrm{QQq}$ can screen at a certain distance. $\mathrm{QQq}$ can not be found anymore above the dashed line. As a comparison, we show  $\rm Q\bar{Q}$ in the same figure with a blue line. The situation is different for $\rm Q\bar{Q}$. The $\rm Q\bar{Q}$ is confined below the blue line. Above the blue line, $\rm Q\bar{Q}$ is deconfined; however, we can still find the $\rm Q\bar{Q}$ at any conditions since it can still exist at a small separation distance.

\section{Conclusion}\label{sec:04}

In this paper, we mainly discuss the properties of $\mathrm{QQq}$ in some aspects at finite temperature and rapidity using the 5-dimensional effective string model for the first time. When the $\mathrm{QQq}$ is in the confined state, it can undergo string breaking at a large separation distance. We consider the breaking mode to be $\mathrm{QQq} \longrightarrow \mathrm{Qqq}+\mathrm{Q\Bar{q}}$ and observe that higher rapidities make the string breaking more difficult to occur. Under high temperature and rapidity conditions, the $\mathrm{QQq}$ transitions into the deconfined state. In this state, it becomes the free quarks if the separation distance is larger than the screening distance, and color screening becomes easier with increasing rapidity. Using potential analysis, we constructed a state diagram for $\mathrm{QQq}$ in the $T-\eta$ plane to differentiate between the confined and deconfined states.

We compared the properties of $\mathrm{QQq}$ and $\mathrm{Q\Bar{Q}}$. As expected, the potential energy of $\mathrm{QQq}$ is always greater than that of $\mathrm{Q\Bar{Q}}$, and the qualitative effects of temperature and rapidity on $\mathrm{QQq}$ and $\mathrm{Q\Bar{Q}}$ are consistent. In the confined state, it is found that the string-breaking distance of $\mathrm{QQq}$ and $\mathrm{Q\Bar{Q}}$ are very close. In other words, it may show the size of the different hadrons are similar. In the deconfined state, there is a significant difference in the screening distance between $\mathrm{QQq}$ and $\mathrm{Q\Bar{Q}}$. It is found that the screening distance of $\mathrm{QQq}$ is smaller than that of $\mathrm{Q\Bar{Q}}$ under the same conditions, which indicates $\mathrm{QQq}$ is less stable than $\mathrm{Q\Bar{Q}}$. Besides, an interesting result shows that $\mathrm{QQq}$ cannot be found above certain temperatures and rapidities. However, $\mathrm{Q\Bar{Q}}$ can be found at any temperature and rapidity as long as the separation distances are small enough. This work mainly focuses on qualitative analysis and can be extended to yield more accurate results in the future.

\acknowledgments

This work is supported by the Natural Science Foundation of Hunan Province of China under
Grant No. 2022JJ40344, and the Research Foundation of Education Bureau of Hunan
Province, China (Grant No. 21B0402).

\end{document}